# Comparing Three Jet Rates With and Without Hadronic Rindler Horizon

Tooraj Ghaffary
Department of Science, Shiraz Branch, Islamic Azad University, Shiraz, Iran
E.mail: ghaffary@iaushiraz.ac.ir

**Abstract**

  Recently, some researchers, (A. Sepehri, S. Shoorvazi, CHIN. PHYS. LETT. Vol. 30, No. 2 (2013)), have considered the effect of Rindler horizon on three jet rate. This paper confirms their results and  by comparing usual models with this new model for different energies, shows that regarding Rindler horizon gives us the results which more close to experimental data respect to usual models.

**Keyword:** $e^+e^-$ annihilation; three jet rate; hadronic Rindler horizon; QCD

## Introduction

The data from $e^+e^-$ annihilation provide us with one of the cleanest ways of probing our quantitative understanding of QCD giving us an opportunity to investigate QCD over the wide range, from measuring the strong coupling between quarks and gluons in low center-of-mass energies to hadronic Rindler horizon in TeV energies. Three-jet production cross sections in $e^+e^-$ annihilation processes are classic hadronic observables which can be measured very accurately and provide an ideal proving ground for testing our understanding of strong interactions and also the temperature TQ of QCD Hawking-Unruh radiation. Three-jet production at tree-level is induced by the decay of a virtual photon (or other neutral gauge boson) into a quark-antiquark-gluon final state. At higher orders, this process receives corrections from extra real or virtual particles. The individual partonic channels that contribute through to NNLO are cited in ref. [1].

In this paper we separate two and three jets events by using the jet clustering algorithm introduced by the JADE group [2,3]. Using computed NNLO corrections to three jet rates, we perform extraction of $\alpha_s$ from data on the standard set of three jet observable measured by the OPAL[4,5] collaborations at center-of-mass energies of 91 GeV to 197 GeV. Then we obtain three jet cross section via ($e^+e^- \rightarrow$ hadronic Rindler horizen $\rightarrow q\bar{q}g$) process and at the end we will compare these cross sections [6].

**JADE algorithm**

We separate two and three jet events by employing the jet clustering algorithm introduced by the JADE group [2]. In this algorithm the scaled mass spread defined as $Y_{ij} = \frac{m_{ij}^2}{E_{vis}^2}$ with $m_{ij}^2 = 2E_iE_j(1-\cos\theta_{ij})$ is calculated for each pair of particles in the event. If the smallest of the $Y_{ij}$ values is less than a parameter $Y_{cut}$, the corresponding pair of particles is combined into a cluster by summing the four momenta. This process is repeated, using all combinations of clusters and remaining particles, until all the $Y_{ij}$ values exceed $Y_{cut}$. The clusters remaining at this stage are defined as the jets.

The distribution of jet multiplicities obtained by these clustering algorithms depends on the jet defining parameter $Y_{cut}$. For small $Y_{cut}$, many jets are found because of the hadronization of fluctuation process, whereas for large $Y_{cut}$, mostly two jet events are found and the $q\bar{q}g$ -events are not resolved. However, Monte Carlo studies show that there is a range of cluster parameters, for which QCD effects can be resolved and the fragmentation effects are sufficiently small. In the following, the parameter $Y_{cut}$=0.04 is used which is found to be a reasonable cut [2].

In figure (1), we show 3-jet fraction for different $Y_{cut}$. The decrease of the 3-jet rate at large $Y_{cut}$ is clearly visible. Our results are completely consistent with the results obtained by the JADE scheme [2,3].

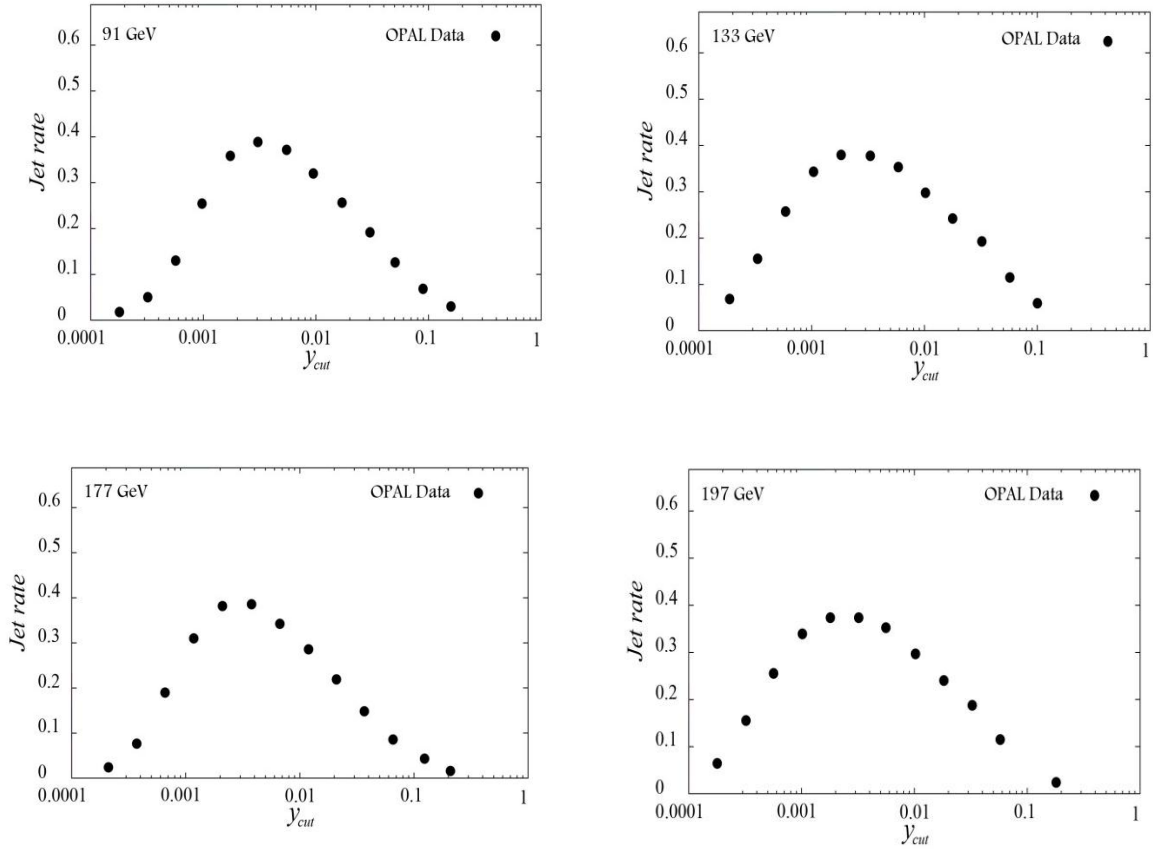

Fig 1. Three jet fraction for different $Y_{cut}$. The data are given from refs[4,5,7].

**Three jet observables at NNLO**

Up to now, the precision of the strong coupling constant determined from three jet observables' data has been limited largely by the scale uncertainty of the perturbative NLO calculation. We report here on the first calculation of NNLO corrections to the 3-jet cross section.

The calculation of the $\alpha_s^3$ corrections for three jet production is carried out using a recently developed parton-level event generator program EERAD3 [8] which contains the relevant matrix elements with up to five external partons. Besides explicit infrared divergences from the loop integrals, the four-parton and five-parton contributions yield infrared divergent contributions if one or two of the final state partons become collinear or soft. In order to extract these infrared divergences and combine them with the virtual corrections, the antenna subtraction method [9] was extended to NNLO level [10] .The three-jet cross section is expanded as [11]

$$\frac{\sigma_{3-jet}}{\sigma_{tot}} = \frac{\alpha_s}{2\pi} A_{3-jet} + (\frac{\alpha_s}{2\pi})^2 B_{3-jet} + (\frac{\alpha_s}{2\pi})^3 C_{3-jet} \qquad (1)$$

The coefficients $A_{3-jet}$, $B_{3-jet}$ and $C_{3-jet}$ are calculated for $\mu^2 = Q^2$ [11]. The measured three-jet cross-sections are shown in graphical form in figure 2 for OPAL Data.

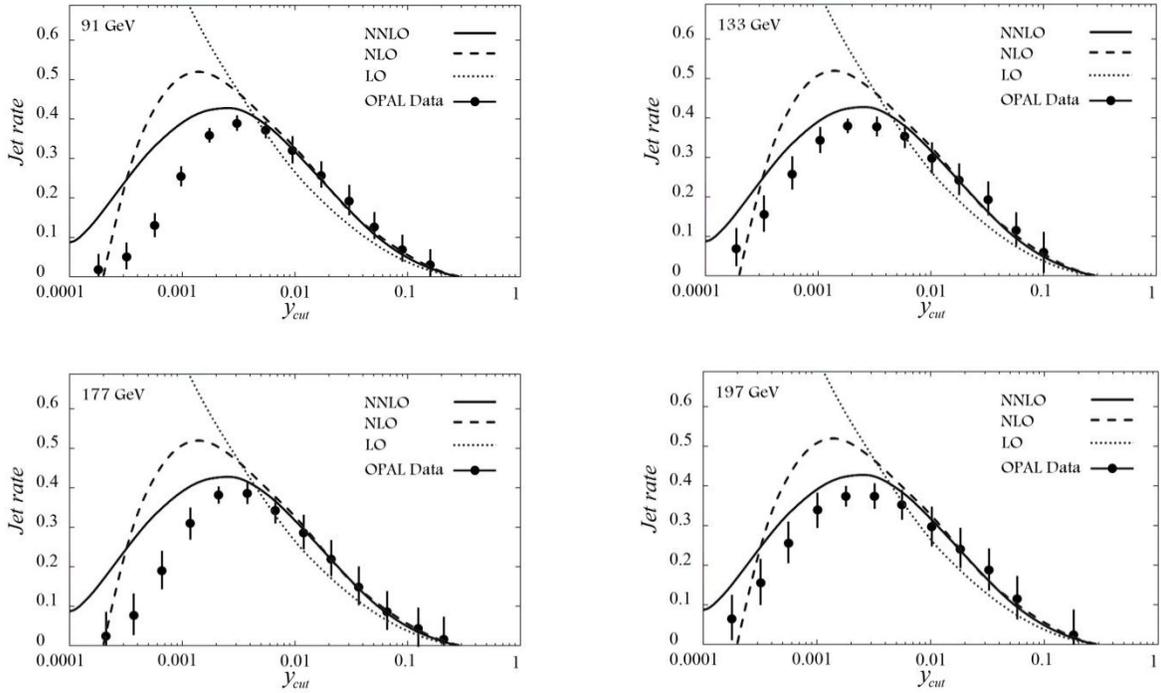

Fig 2. The scale variation of the three jet rate with the JADE jet algorithm. The data are given from refs [4,5,7]and model is given form ref [11].

By comparing the three plots, on each figure we observe that the agreement for each of the jet rates becomes systematically better as the order of perturbation theory increases. By fitting our data with equation (1) the strong coupling constant $α_s$ can be derived. The mean value of $α_s$ from these calculations is 0.116±0.004.

**Three jet cross section via $e^+e^- →$ hadronic Rindler horizen $→ q\bar{q}g$**

Until now, we have considered usual three jet rate cross section. However we regard one Rindler horizon that is produced by electron-positron annihilation and then decay to particles and formed the three jet rate cross section. This method is as follows [6].

The investigation of three jet cross section through $e^+e^- →$ hadronic Rindler horizen $→ q\bar{q}g$ process consists first in duplicating the degrees of freedom of the system. For this purpose, a copy of the Hilbert space for QCD can be constructed with a set of operators of creation/annihilation that have the same commutation properties as the original ones. The whole Hilbert space is the tensor product of the two spaces, which in this case explains the physical quantum states space of the quarks-antiquarks inside and outside the hadronic Rindler horizon. The quarks which are lying outside the hadronic Rindler horizon impress the antiquarks inside the horizon by confinement. Thus we can rewrite the Bogoliubov transformation description between the Minkowski and black hole creation and annihilation operators:

$$(q - \tanh r_\omega \bar{q})|hadronic\ Rindler\ horizon\rangle_{inside \otimes outside} = 0 \qquad (2)$$

Where

$$\tanh r_\omega = e^{-\frac{4\pi\omega}{T_Q}} \qquad (3)$$

which actually organize a boundary state. In this equation QCD temperature can be get as follow [12]:

$$T_Q = \sqrt{\frac{m^2}{4\pi\alpha_s}} \tag{4}$$

where $\alpha_s$ is the strong coupling constant and m is quark mass. By the expansion of equation (2) in modes of quarks and antiquarks we have:

$$(\alpha_q - \tanh r_\omega \alpha_{\bar{q}})|hadronic\ Rindler\ horizon\rangle_{inside \otimes outside} = 0 \tag{5}$$

Now, we can see that the relevance between hadronic Rindler vacuum $|hadronic\ Rindler\ horizon\rangle_{inside \otimes outside}$ and the confinement vacuum $|0\rangle_{confinment}$ is

$$|hadronic\ Rindler\ horizon\rangle_{inside \otimes outside} = F|0\rangle_{confinment} \tag{6}$$

here F is the function which will be defined later.

According to $\{\alpha_q, \alpha_q^\dagger\} = 1$, we will have $\{\alpha_q, (\alpha_q^\dagger)^m\} = \frac{\partial}{\partial \alpha_q^\dagger}(\alpha_q^\dagger)^m$ and $[\alpha_q, F] = \frac{\partial}{\partial \alpha_q^\dagger} F$. Due to equation (5) and (6), we obtain the subsequent differential equation for F:

$$\left(\frac{\partial F}{\partial \alpha_q^\dagger} - \tanh r_\omega\ \alpha_q^\dagger F\right) = 0 \tag{7}$$

The solution of equation (7) is:

$$F = e^{\tanh r_\omega \alpha_q^\dagger \alpha_{\bar{q}}^\dagger} \tag{8}$$

Using equation (8) and equation (6) and by normalizing the state vector, we have [6]:

$$|hadronic\ Rindler\ horizon\rangle_{inside \otimes outside} = Ne^{\tanh r_\omega \alpha_q^\dagger \alpha_{\bar{q}}^\dagger}|0\rangle_{confinment}$$
$$= \frac{1}{\cosh r_\omega}\sum_m \tanh^m r_\omega |m\rangle_{inside} \otimes |\bar{m}\rangle_{outside} \tag{9}$$

where $|m\rangle_{inside}$ and $|\bar{m}\rangle_{outside}$ are quarks and antiquarks orthonormal bases that are operated on H$_{inside}$ and H$_{outside}$ respectively and N is the normalization constant.

Equation (9) explains that the confinement state decomposes to many entangled states with different energies at boundary. From other view, this equation states that non-local physics would be needed to send the information inside the hadronic Rindler horizon and also the Hilbert spaces inside and outside the hadronic Rindler horizon do not have independent existence.

The quarks and antiquarks' thermal distribution will be obtained as follow:

$$N_\omega = {}_{inside \otimes outside}\langle hadronic\ Rindler\ horizon|\alpha_q^\dagger \alpha_q|hadronic\ Rindler\ horizon\rangle_{inside \otimes outside} = \frac{e^{\frac{-2\pi\omega}{T_Q}}}{1-e^{\frac{-2\pi\omega}{T_Q}}} \qquad (10)$$

Equation (10) demonstrates that different amount of quarks and antiquarks produced with different probabilities inside and outside the hadronic Rindler horizon which these probabilities are related to their energies. In this part, we will get the radius of the hadronic Rindler horizon as follow [12]:

$$R_Q = \frac{2\alpha_s}{m} \qquad (11)$$

To get the total three jet cross section, we should multiply the hadronic Rindler horizon production cross section by the number of quarks and antiquarks produced near this horizon by the probability for radiation gluon from quark or antiquark.

$$\sigma^{three\ jet} = \int_{Q'}^{Q_0} \frac{dq}{q^2} \int_m^{Q'} d\omega \int_m^{Q'} d\omega' \int_{\frac{q}{\omega}}^{1-\frac{q}{\omega}} dz\ N_\omega^{quark} N_{\omega'}^{antiquark}\ \Gamma_{q \to qg}\ \sigma^{hadronic\ Rindler\ horizon} \qquad (12)$$

where $Q'$ is the jet resolution scale in this model, $(Q' = y_{cut}Q_0)$, $Q_0$ is the center-of-mass energy of the colliding $e^+e^-$ and q is considered as the gluon momentum.

Considering the black hole production cross section at LHC [13,14], the production cross section for hadronic Rindler horizon is being calculated:

$$\sigma^{\text{hadronic Rindler horizon}} = \pi R_Q^2 \tag{13}$$

Meanwhile we calculate the probability for radiation gluon from quark or antiquark in $D = 4 - 2\epsilon$ as follow [15,16]:

$$\Gamma_{q \to qg}(z) = \frac{4}{3}\left[\frac{1+z^2}{1-z} - \epsilon(1-z)\right] \tag{14}$$

where z is the fraction of quark energy that carries by gluon. We approximate the integral in equation (12) as [6]:

$$\sigma^{\text{three jet}} = \left[\ln\left(\frac{m}{y_{cut}Q_0}\right) + \ln\left(\frac{16\pi y_{cut}Q_0 - \frac{3}{2\alpha_s}}{16\pi m - \frac{3}{2\alpha_s}}\right)\right]\alpha_s^3 \left[\frac{32}{27\pi}\left[\ln\left(\frac{m}{y_{cut}}\right) + y_{cut}Q_0\ln\left(\frac{1}{y_{cut}}\right) + Q_0(1-y_{cut})\left(\ln\left(\frac{1-y_{cut}}{y_{cut}}\right) + 2(1-y_{cut})\right)\right]\right] + \alpha_s^4 \left[\frac{128}{810}\left[\left(\left(\frac{y_{cut}Q_0}{2}\right)^2 - \frac{m^2}{2}\right)(1-y_{cut})\ln\left(\frac{1}{y_{cut}}\right) + \left(\left(\frac{\ln^3\left(\frac{Q_0}{y_{cut}}\right)}{3}\right) - \frac{m^3}{3}\right)(1-y_{cut}) + \left(\left(\frac{y_{cut}Q_0}{2}\right)^2 - \frac{m^2}{2}\right)(1-y_{cut})\right)\right]\right] \tag{15}$$

Equation (15) shows that by increasing the center-of-mass energy, the effect of QCD horizon on three jet cross section increases. We can see, as the order of perturbation theory increases, the effect of hadronic horizon on hadronic cross section becomes systematically more effective. This is because at higher orders, there exists more channels for QCD horizon production and it's decays into massive quark-anti quarks in our calculation.

**Comparing the hadronic Rindler horizon model with experimental data**

In figure (3) we have compared three jet cross section via $e^+e^- \to$ hadronic Rindler horizen $\to q\bar{q}g$ process with experimental data from

OPAL Collaboration[16]. In this plot we take m=5GeV for quark mass, $\alpha_s = 0.11$ and $Q_0 = 91$ to 197GeV for center-of-mass energies. Also, in this figure we add the JETSET/PYTHIA and HERWIG Monte Carlo expectations. As can be seen from this figure, the three jet cross sections is rising at $y_{cut} = 0.0001$, represent a turn-over at moderate value of $y_{cut} = 0.003$ to $0.005$ and then by increasing the $y_{cut}$, it decreases rapidly. This model is in good agreement with OPAL data.

We come to this end that the process of electron -positron annihilation at center-of-mass energies of GeV, can create a hadronic Rindler horizon and we can expect that this horizon is a QCD matter factory. Near QCD horizon, We are treating the quarks and gluons as free particles. This is because their energies are high and they are produced inside and outside of hadronic Rindler horizon and don't access to each other. So it's not expected that they hadronized until they radiate gluons and forming three jet events.

As can beseen, it is clear that the cross section of three jet events produced via gluon radiation near a single Hadronic Rindler Horizon is much larger for higher center of mass energies, because as the energy of the Hadronic Rindler Horizon becomes higher, the temperature becomes larger and the thermal radiation of gluons is enhanced.

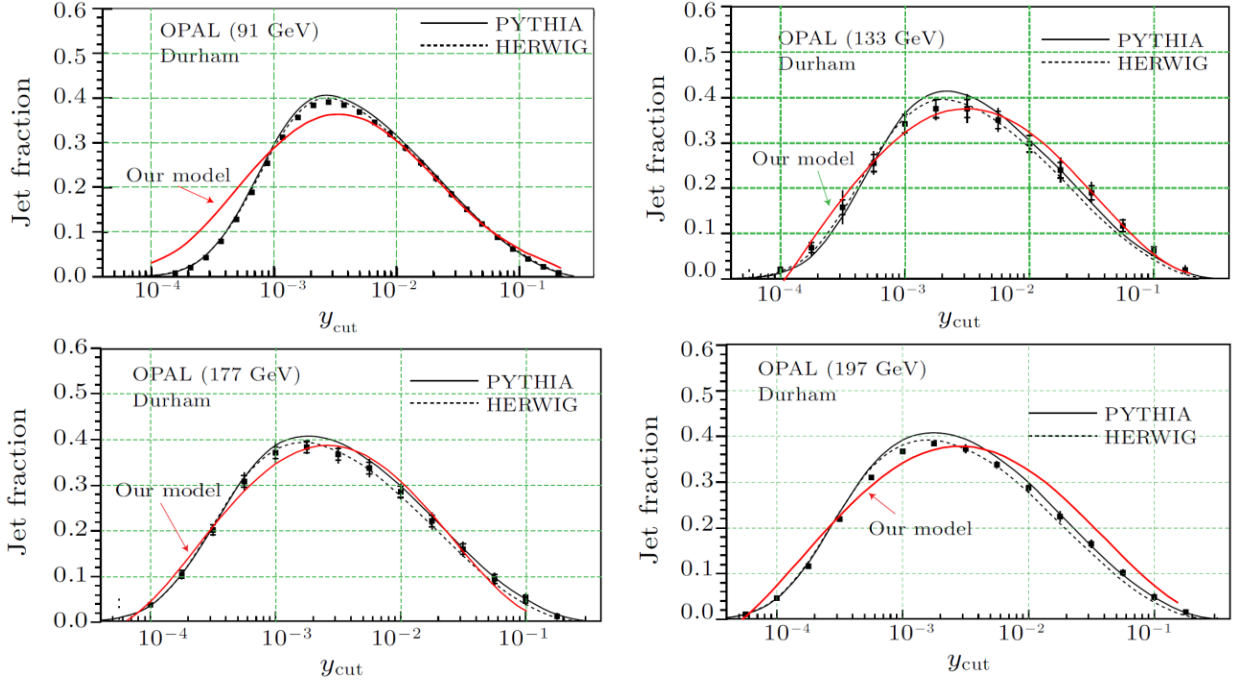

Fig.3: A three-jet cross section via the $e^+e^- \to Hadronic\ Rindler\ Horizon \to q\bar{q}g$ process compared to the experimental data from 91 to 197 GeV [6,17].

**Conclusion**

When electron-positron annihilate, different events may be occurred. Until now, it has been argued that by disappearing these particles at energies beyond GeV, only quark-anti-quark and gluon are produced that form three jet rates, four jet rates and other jet rate cross section. Recently, some authors, (A. Sepehri, S. Shoorvazi, CHIN. PHYS. LETT. Vol. 30, No. 2 (2013)), have consider the effect of Rindler horizon on three jet rate and show that regarding this horizon give the better results. In this research, the experimental data from 91 to 197 GeV are compared with models with and without Rindler horizon and show that this horizon gives us the exact results.